\documentclass[10pt,english,conference]{IEEEtran}
\usepackage{mathrsfs}
\usepackage{amsfonts}
\usepackage{amsthm}
\usepackage{graphicx,cite,epsfig,amssymb,amsmath}
\usepackage{color,xcolor}
\usepackage{pifont}
\usepackage{stmaryrd}
\usepackage{bbding}
\usepackage{setspace}
\usepackage{indentfirst}
\usepackage{subfigure}
\usepackage[hypertex]{hyperref}
\usepackage{cite}

\usepackage{algorithmic,algorithm}

\newcommand{\tabincell}[2]{\begin{tabular}{@{}#1@{}}#2\end{tabular}}

\begin{document}

\title{Spatial Resource Allocation for Spectrum Reuse in Unlicensed LTE Systems}

\author{\authorblockN{Rui Yin$^{1,2}$, Amine Maaref$^3$, and Geoffrey Ye Li$^4$
\thanks{This work was supported in part by National Natural Science Foundation Program of China under Grant No. 61771429, Natural Science Foundation Program of Zhejiang Province under Grant No. LY14F010002, and Natural Science and Technology Specific Major Projects No. 2015ZX03001031-003.}}

\authorblockA{1. School of Information and Electrical Engineering, Zhejiang University City College, Hangzhou, China\\
 2. School of Electrical, Electronic and Computer Engineering, University of KwaZulu-Natal, Durban, South Africa\\
 3. Canada Research Centre, Huawei Technologies Canada Co., Ltd, Ottawa, Ontario, Canada\\
 4. School of Electrical and Computer Engineering, Georgia Institute of Technology, United States\\
 Email: xrk2000@gmail.com, amine.maaref@huawei.com, liye@ece.gatech.edu}}
\maketitle
\begin{abstract}
In this paper, we study how to reuse the unlicensed spectrum in LTE-U systems while guaranteeing harmonious coexistence between the LTE-U and Wi-Fi systems. For a small cell with multiple antennas at the \emph{base station} (SBS), some spatial \emph{degrees of freedom} (DoFs) are used to serve \emph{small cell users} (SUEs) while the rest are employed to mitigate the interference to the Wi-Fi users by applying \emph{zero-forcing beamforming} (ZFBF). As a result, the LTE-U and Wi-Fi throughput can be balanced by carefully allocating the spatial DoFs.
Due to the \emph{channel state information} (CSI) estimation and feedback errors, ZFBF cannot eliminate the interference completely. We first analyze the residual interference among SUEs, called intra-RAT interference, and the interference to the Wi-Fi users, called inter-RAT interference after ZFBF, due to imperfect CSI. Based on the analysis, we derive the throughputs of the small cell and the Wi-Fi systems, respectively. Accordingly, a spatial DoF allocation scheme is proposed to balance the throughput between the small cell and the Wi-Fi systems. Our theoretical analysis and the proposed scheme are verified by simulation results.
\end{abstract}

\section{Introduction}
The scarce licensed spectrum is the main bottleneck to further improve the data rates in \emph{the fifth generation} (5G) wireless communications. To alleviate this spectrum gridlock, \emph{spectrum efficiency} (SE) on the licensed bands has been improved at the greatest extent. On the other hand, extra spectrum resource on the unlicensed bands has been also considered for LTE, which is called LTE-U \cite{14_Huawei}.

Most current works focus on how to improve the SE on the licensed bands in LTE systems. By exploiting the spatial freedom, \emph{multiple-input multiple-output} (MIMO) can boost the SE by the scale of the number of antennas \cite{14_v}, especially in the ultra-dense \emph{small cell base station} (SBS) on the unlicensed spectrum with both \emph{small cell users} (SUEs) and Wi-Fi users.

Conventionally, unlicensed bands are shared by different \emph{radio access technologies} (RATs) and mainly occupied by the Wi-Fi systems. In order to share the unlicensed bands harmoniously, distributed channel access is used in Wi-Fi \cite{13_Eldad}. In contrast, LTE is with centralized-scheduling structure due to its exclusive usage of the licensed bands \cite{14_LTE} and has higher SE. How to exploit highly efficient spectrum management techniques of LTE in the unlicensed bands and how to guarantee the harmonious coexistence with Wi-Fi systems are the most important issues to realize an LTE-U system.

With little modification on the current LTE protocol, a duty cycle method has been proposed for LTE-U in \cite{13_Alme}, where LTE periodically turns the signal on and off by using \emph{almost blank subframe} (ABSs) on the unlicensed bands and the Wi-Fi users access the unlicensed bands when the LTE signal is off. To further reduce the impact to the Wi-Fi system, \emph{listen-before-talk} (LBT) mechanism has been proposed for the LTE-U systems in \cite{12_ETSI}, where LTE senses the unlicensed bands before using them. In \cite{16_yin}, we have proposed an adaptive channel access scheme based on LBT mechanism for LTE-U systems. When a SBS shares unlicensed channels with the Wi-Fi system and reuses the licensed channels with the macro cell, the tradeoff between the collision probability experienced by the Wi-Fi users and the co-channel interference to the macro cell users has been analysed in \cite{16_yin2}.

In the paper, we study how to use MIMO in the LTE-U systems to reuse the unlicensed spectrum. To guarantee a fair coexistence between the LTE-U and the Wi-Fi systems, \emph{zero-forcing beamforming} (ZFBF) is employed at the SBS to mitigate the interference among SUEs, called intra-RAT interference, and the interference to the Wi-Fi system, called
inter-RAT interference. First, a general framework to realize the unlicensed spectrum reuse by applying ZFBF technique is formulated for the LTE-U system. Since the \emph{channel state information} (CSI) available at the SBS is imperfect, the residual
intra-RAT and inter-RAT interference are analyzed first. Then, the achievable throughputs for the small cell and the Wi-Fi are derived. Since throughputs of the small cell and the Wi-Fi depend on the assigned spatial \emph{degree of freedom} (DoF), we will develop a DoF allocation scheme to balance the small cell and Wi-Fi throughput.



\section{System Model}\label{s2}
In the paper, we assume that an LTE-U SBS coexists with one Wi-Fi \emph{access point} (AP) to share one unlicensed band simultaneously in the system. The SBS serves downlink transmission to SUE set, $\mathcal{U}$, with $|\mathcal{U}|=U$, and the Wi-Fi AP serves user set, $\mathcal{M}$, with $|\mathcal{M}|=M$. We assume that the Wi-Fi users have the same type of data traffic. Therefore, the number of the Wi-Fi users can represent the Wi-Fi traffic load. The SUE and the Wi-Fi users have one antenna. The SBS has $N_T$ antennas.
ZFBF is employed at the SBS to mitigate the intra-RAT interference among SUEs, and the inter-RAT interference to the Wi-Fi users during the downlink transmission. It also mitigates the inter-RAT interference to the Wi-Fi AP during the Wi-Fi uplink transmission.
If $N$ spatial DoFs are used to serve $K$ SUEs, denoted as a set $\mathcal{U}_s$ ($K=|\mathcal{U}_s|$), in the small cell system and one DoF is used to mitigate interference for the Wi-Fi AP, then the left over spatial DoFs are used to mitigate the interference to $N_T-N-1$ Wi-Fi users, denoted as a set $\mathcal{M}_s$ ($N_T-N-1=|\mathcal{M}_s|$).
On the Wi-Fi side, Wi-Fi users compete for the unlicensed band based on the DCF mechanism where an exponential backoff scheme with the minimum backoff window size, $W$, and maximum contention stage, $L$, are employed to avoid the collision.

Denote $\boldsymbol{h}_k$ as the channel vector between the SBS and the SUE $k$. The elements in the channel vector are assumed to be \emph{independent and identical distributed} (i.i.d.) complex Gaussian random variables with zero mean and unit variance, which can be obtained at the SBS via the feedback from the SUEs. To mitigate the interference to the Wi-Fi users, the SBS also needs to know the CSI to the Wi-Fi users and to the Wi-Fi AP. We assume that the Wi-Fi AP interface is integrated into the SBS \cite{16_chen}. Therefore the SBS can monitor the Wi-Fi signals on the unlicensed spectrum. After receiving the preambles from the Wi-Fi users and the AP, the SBS can estimate CSI from the Wi-Fi users and the Wi-Fi AP to the SBS. The CSI from the Wi-Fi users or the Wi-Fi AP to the SBS can be obtained by exploiting channel reciprocity

\subsection{SINR of small cell users}
Denote $A_m$ as the path-loss factor of the interference channel from the SBS to the Wi-Fi user $m$,  ${{\bf{f}}_m}$ as the corresponding small-scale channel fading vector from the SBS to the Wi-Fi user $m$, and ${\bf{D}}$ as the channel fading vector from the SBS to the Wi-Fi AP. All elements in the above defined vectors are i.i.d. complex Gaussian random variables with zero mean and unit variance. Then, with the overall CSI, for SUE $k\in\mathcal{U}$, its complementary channel matrix is
given by
\begin{equation}\label{eq1}
\small
{\overline H _k} = [{{\bf{f}}_1}, \cdots ,{{\bf{f}}_{N_T-N-1}},\bf{D}, {{\bf{h}}_1}, \cdots ,{{\bf{h}}_{k - 1}},{{\bf{h}}_{k + 1}}, \cdots, {{\bf{h}}_K}].
\end{equation}
\noindent For ZFBF, the precoder, ${\bf{v}}_k$, for SUE $k$ can be derived based on the \emph{singular value decomposition} (SVD) to its complementary channel matrix $\overline H _k$. Denote $U_k^{\perp}$ as the null space of $\overline H _k$, which is spanned by the right singular vectors with zero singular value. The precoder, ${\bf{v}}_k$, can be calculated as the normalized projection vector from $\bf{h}_k$ to $U_k^{\perp}$. With ZFBF, the received signal at SUE $k$ is given by
\begin{eqnarray}\label{eq1_2}
\small
{y_k}&=& \sqrt {\frac{{{P_T}}}{K}} {\bf{h}}_k^H{{\bf{v}}_k}{s_k} + \sqrt {\frac{{{P_T}}}{K}} \sum\limits_{i = 1,i \ne k}^K {{\bf{h}}_k^H{{\bf{v}}_i}{s_i}}  + {n_k},
\end{eqnarray}
\noindent where $P_T$ is the transmit power, $s_k$ is the transmitted signal to SUE $k$,
$n_k$ is sum of the noise and interference received at SUE $k$.
If accurate CSI is known at the SBS, then the intra- and inter-RAT interference can be eliminated completely, e.g., ${\bf{h}}_i^H{\bf{v}}_k=0,~\forall i\ne k,~i\in\mathcal{U}_s,~{\bf{f}}_m^H{\bf{v}}_k=0,~\forall m\in \mathcal{M}_s,~{\bf{D}}^H{\bf{v}}_k=0$.

However, in practice, the SBS obtains the CSI from the quantization codeword index feedback from the SUEs \cite{09_zhang}.
According to
the random vector quantization theory \cite{07_Isu}, the normalized CSI, ${{\tilde{\bf{h}}}_k}={\bf{h}}_k/\left\| {{{\bf{h}}_k}} \right\|$, for SUE $k$ can be decomposed into
\begin{equation}\label{quant}
\small
{\widetilde{\bf{h}}_k} = \sqrt {1 - b} {\widehat{\bf{h}}_k} + \sqrt b {\bf{c}},
\end{equation}
where $b$ indicates the quantization error, scaling from 0 to 1, $\widehat{\bf{h}}_k$ is the estimated channel vector obtained by feedback and $\bf{c}$ is a unit norm vector isotropically distributed in the null space of $\widehat{\bf{h}}_k$ and is independent of $b$. If $b=0$, then the SBS has the perfect CSI. $b$ is related to the codebook size $2^B$ and is upper bounded by  $2^{-\frac{B}{N_t-1}}$ \cite{06_jindal}.
Based on \eqref{quant}, \eqref{eq1_2} can be rewritten as
\begin{equation}\label{quant2}
\small
{y_k} = \sqrt {\frac{{{P_T}}}{K}} {\bf{h}}_k^H{{\bf{v}}_k}{s_k} + \sqrt {\frac{{{P_T}}}{K}} \sum\limits_{i = 1,i \ne k}^K {\sqrt{b}{{\left\| {{{\bf{h}}_k}} \right\|}^2}{{\bf{c}}^H}{{\bf{v}}_i}{s_i}}  + {n_k},
\end{equation}
where $\sqrt {\frac{{{P_T}}}{K}} \sum\nolimits_{i = 1,i \ne k}^K {\sqrt{b}{{\left\| {{{\bf{h}}_k}} \right\|}^2}{{\bf{c}}^H}{{\bf{v}}_i}{s_i}}$ is the residual interference due to the error in available channel vector since $\widehat{\bf{h}}_k^H{{\bf{v}}_i} = 0$ for $\forall i\neq k$.
Based on equation \eqref{quant2}, the achievable \emph{signal-to-interference-plus-noise power ratio} (SINR) at SUE $k$ is given by
\begin{equation}\label{quant3}
\small
SINR_k=\frac{{{{\left| {{\bf{h}}_k^H{{\bf{v}}_k}} \right|}^2}}}{{\frac{K}{{{P_T}}}N_0 + b{{\left\| {{{\bf{h}}_k}} \right\|}^2}\sum\limits_{i = 1,i \ne k}^K {{{\left| {{{\bf{c}}^H}{{\bf{v}}_i}} \right|}^2}} }},
\end{equation}
where $N_0$ is the addition of noise and interference power. Notice that the interference from the Wi-Fi users to the SUEs is treated as the noise in the paper. From \eqref{quant3}, we can observe that the achievable SINR in SUE $k$ is restricted by the
residual interference due to the CSI error.

\subsection{Wi-Fi throughput}
Based on \cite{00_bian}, the Wi-Fi user transmission probability on the unlicensed band can be written as
\begin{equation}\label{access}
\small
\tau  = \frac{{2 \times (1 - 2{p_F})}}{{(1 - 2{p_F})(W + 1) + {p_F}W(1 - {{(2{p_F})}^L})}},
\end{equation}
where
\begin{equation}\label{collision}
\small
{p_F} = 1 - {(1 - \tau )^{M - 1}}
\end{equation}
is the collision probability and
\begin{equation}\label{succes}
\small
P_S=\frac{M\tau(1-\tau)^{M-1}}{1-(1-\tau)^M}
\end{equation}
is the conditional successful transmission probability for the Wi-Fi users.

Based on \eqref{succes}, the Wi-Fi throughput can be written as \cite{00_bian}
\begin{equation}\label{wf_throughput}
\small
R_w = \frac{{{P_T}{P_S}\mathbf{E}\{ Package\} }}{{(1 - {P_T})\delta  + {P_T}{P_S}{Q_s} + {P_T}(1 - {P_S}){Q_c}}},
\end{equation}
where
\begin{equation}\label{Tran_prob}
\small
P_T=1-(1-\tau)^M,
\end{equation}
is the probability that there is at least one transmission on the unlicensed channel,  $\mathbf{E}\{ Package\}$ is the average packet payload size for Wi-Fi transmission, $\delta$ is the duration of an empty time slot, $Q_s$ is the average channel occupied time due to a successful transmission, and $Q_c$ is the average channel busy time sensed by the Wi-Fi users due to collision~\cite{00_bian}. From \eqref{collision}, \eqref{succes}, \eqref{wf_throughput} and \eqref{Tran_prob}, the Wi-Fi throughput is directly related to the number of the Wi-Fi users, $M$, competing for the unlicensed bands.

When the SBS reuses the unlicensed band, the Wi-Fi users will suffer inter-RAT interference, including those Wi-Fi users selected for interference cancellation by ZFBF at the SBS due to imperfect CSI available at the SBS.
We use the common channel uncertain model \cite{07_Isu} to express the channel vector between the SBS and the Wi-Fi user $m$
\begin{equation}\label{eq3}
\small
{{\bf{f}}_m} = \sqrt \varepsilon  {{\bf{f}}_{m,0}} + \sqrt {1 - \varepsilon } {\boldsymbol{\phi}},
\end{equation}
where ${\bf{f}}_{m,0}$ is the estimated CSI between the SBS and the Wi-Fi user $m$, $\boldsymbol{\phi}$ is the estimation error vector
with zero mean and unit variance complex Gaussian distributed entries. $\varepsilon$ is the correlation factor.
Since the locations of the Wi-Fi AP and the SBS are always static, we assume that the SBS can obtain the perfect CSI from the SBS to the Wi-Fi AP. Therefore, we have $~\boldsymbol{D}^H\boldsymbol{v}_k=0,~\forall k\in{\mathcal{U}_s}$.

\section{Performance Analysis}\label{s3}
In this section we first analyze the intra-RAT residual interference among SUEs and derive the throughput of the small cell on the unlicensed band. Then, inter-RAT interference to the Wi-Fi users are analyzed and the Wi-Fi system throughput is derived accordingly.

\subsection{SBS throguhput}
Based on \eqref{quant3}, the \emph{cumulative distribution probability} (CDF) of $x \sim SIN{R_k}$ can be expressed in the following lemma.

\textbf{Lemma 1:} \emph{The CDF of $SINR_k$ for each SUE $k$ is given by}
\begin{equation}\label{CDF_SUE}
\small
F(x)=1 - \frac{{\exp ( - \frac{K}{{{P_T}}}N_0x)}}{{{{(1 + \sigma x)}^{ K - 1}}}}.
\end{equation}

Denote $f(x)=F'(x)$ as the corresponding \emph{probability density function} (pdf). Then, we can derive the throughput of the small cell as
\begin{eqnarray}\label{SB_th}
\small
 {R_s} &=& K\int_0^\infty  {\log (1 + x)f(x){d_x}} \\
  &=& \frac{{K\log_2(e)}}{{{\sigma ^{K - 1}}}}\psi ( - \frac{{K{N_0}}}{{{P_T}}},{\sigma ^{ - 1}},K - 1),\nonumber
\end{eqnarray}
where
\begin{eqnarray}
\small
 &&\psi (x,y,z)=\int_0^\infty  {\frac{{\exp ( - xt)}}{{(t + 1){{(t + y)}^z}}}{d_t}}\nonumber  \\
  &=&\sum\limits_{i = 1}^z {{( - 1)}^{i - 1}}{{(1 - y)}^{ - i}}{I_2}(x,y,z - i + 1) \nonumber \\
  &&+ {{(y - 1)}^{ - z}}{{I_2}(x,1,1)},\nonumber
\end{eqnarray}
and
\begin{eqnarray}
\left\{ {\begin{array}{*{20}{c}}
   {\exp (xy){E_1}(xy),z = 1,}  \\
   {\sum\limits_{k = 1}^{z - 1} {\frac{{(k - 1)!}}
{{(z - 1)!}}\frac{{{{( - x)}^{z - k - 1}}}}
{{{y^k}}} + \frac{{{{( - x)}^{z - 1}}}}
{{(z - 1)!}}\exp (xy){E_1}(xy),z \geqslant 2,} }  \\
 \end{array} } \right. \nonumber
\end{eqnarray}
where $E_1(\cdot)$ is the exponential integral function of the first order. According to \eqref{SB_th}, the small cell throughput is related to the number of served SUEs, $K$, the feedback bits, B, and the total transmission power, $P_T$.

As we have indicated after
\eqref{quant3}, intra-RAT interference increases with the number of served SUEs. As a result, the system throughput may not increase with $K$. Therefore, there exists an optimal value $K$ to maximize the small cell throughput for given feedback bits, $B$, and total transmission power, $P_T$.

\subsection{Wi-Fi throughput}
Inter-RAT interference to the Wi-Fi users can be divided into two categories according to whether they are selected for interference cancellation. If the Wi-Fi users are selected for the interference cancellation at
the SBS, then there exists residual interference due to imperfect CSI. According to \eqref{eq3}, the residual interference from the SBS to Wi-Fi user $m$ is given by
\begin{eqnarray}\label{eq4}
\small
{I_m} &=& \frac{{{P_T}}}{K}{A_m}\sum\limits_{k = 1}^K {\left| {{\bf{f}}_m^H{{\bf{v}}_k}} \right|^2}  = \frac{{{P_T}}}{K}{A_m}\sum\limits_{k = 1}^K {\left| {\sqrt {1 - \varepsilon } {{\boldsymbol{\phi}}^H}{{\bf{v}}_k}} \right|^2}\nonumber\\
&=& \frac{{{P_T}}}{K}{A_m}(1 - \varepsilon )\sum\limits_{k = 1}^K {{{\left| {{{\boldsymbol{\phi}}^H}{{\bf{v}}_k}} \right|}^2}},
\end{eqnarray}
Then, the pdf of variable $\sum\nolimits_{k = 1}^K {{{\left| {{{\boldsymbol{\phi}}^H}{{\bf{v}}_k}} \right|}^2}}$ is as in the following lemma.

\textbf{Lemma 2:} \emph{The variable $x\sim \sum\limits_{k = 1}^K {{{\left| {{{\boldsymbol{\phi}}^H}{{\bf{v}}_k}} \right|}^2}}$ is Gamma distributed, whose PDF is given by
\begin{equation}\label{ga}
\small
f(x;2K,K) = \frac{{{x^{2K - 1}}{e^{ - \frac{x}{K}}}}}{{{K^{2K}}\Gamma (2K)}},~\rm{for}~x>0,
\end{equation}
where $\Gamma(2K)$ is the Gamma function evaluated at $K$.
}
\begin{proof}
Since elements in $\boldsymbol{\phi}$ are complex Gaussian distributed with zero mean and unit variance in \eqref{eq3}, ${\left| {{{\boldsymbol{\phi}}^H}{{\bf{v}}_k}} \right|}^2$ is exponential distributed with mean 1 \cite{09_zhang}. Then $x = \sum\nolimits_{k = 1}^K {{{\left| {{\phi ^H}{{\bf{v}}_k}} \right|}^2}}$, the sum of exponential distributed variables, is wtih Gamma distribution of parameters $2K$ and $K$, as expressed in \eqref{ga}.
\end{proof}
On the other hand, if the Wi-Fi user is not selected for interference cancellation, inter-RAT interference from the SBS will be high, which is given by
\begin{equation}\label{eq41}
\small
{\widehat{I}_n} = \frac{{{P_T}}}{K}{A_n}\sum\limits_{k = 1}^K {{{\left| {{\bf{f}}_n^H{{\bf{v}}_k}} \right|}^2}},~\forall n\notin\mathcal{M}_s,~n\in\mathcal{M}.
\end{equation}
From Lemma 2, the variable $\sum\limits_{k = 1}^K {{{\left| {{\bf{f}}_n^H{{\bf{v}}_k}} \right|}^2}}$
is also Gamma distributed with the same pdf as \eqref{ga}.

If the interference power experienced by a Wi-Fi user is greater than a threshold $\bar{I}$, i.e. ${I_m} > \bar{I}~{\rm{or}}~{\widehat{I}_n} > \bar{I}$, the Wi-Fi user is unable to access the unlicensed band. Therefore, the access probability for selected Wi-Fi user $m$ and the non-selected Wi-Fi user $n$ is written as $P_r(m)=P_r(I_m\le\bar{I})$ and $P_r(n)=P_r(\widehat{I}_n\le\bar{I})$, respectively.
With the access probability, the average number of Wi-Fi users competing for the access to the unlicensed channel is $\overline M = \sum\nolimits_{i\in\mathcal{M}} {{P_r(i)}}$.
Then, substituting $\overline{M}$ into \eqref{access},~\eqref{collision}, and~\eqref{succes}, we can achieve the Wi-Fi system throughput from \eqref{wf_throughput}. Based on
\cite{00_bian}, the Wi-Fi throughput increases with the number of the active Wi-Fi users if the Wi-Fi system is with light load. On the other hand, it decreases with the number of the active Wi-Fi users if the Wi-Fi system is overloaded. Therefore, the Wi-Fi system throughput is convex with respect to the number of the Wi-Fi users. In this paper, we focus on the scenario that the Wi-Fi system is not overloaded.

\section{DoF Allocation}
The tradeoff between the SBS and the Wi-Fi system throughput can be achieved via spatial DoF allocation based on the performance analysis in the last section.
Obviously, more spatial DoFs used for the inter-RAT interference mitigation, less SUEs can be served by the SBS. As a result, more Wi-Fi throughput can be achieved. On the other hand, if fewer spatial DoFs are applied to mitigate the inter-RAT interference to the Wi-Fi users, more spatial DoFs are available at the SBS and higher SBS throughput is expected.
Therefore, there is a fundamental tradeoff between the small cell and the Wi-Fi throughput via spatial DoF allocation. To analyze this tradeoff while guaranteeing the fair coexistence,
We formulate the problem as
\begin{eqnarray}\label{op1}
\small
\mathop {\max }\limits_{{\mathcal{U}_s},{\mathcal{M}_s},N}\min\{ {e_sR_s}, {e_wR_w}\}
\end{eqnarray}
subject to
\begin{align}
\small
&|\mathcal{U}_s|=K,~\frac{R_s}{K}\ge \bar{r}_s,\label{op1_c2}\tag{\theequation b}\\
&\frac{R_w}{\overline{M}}\ge \bar{r}_w,\label{op1_c3}\tag{\theequation c}\\
&2\le K\le N,\label{op1_c4},\tag{\theequation d}\\
&N+|\mathcal{M}_s|= N_t-1, \label{op1_c5}\tag{\theequation e}
\end{align}
where $e_s$ and $e_w$ are weight factors associated to the SBS and Wi-Fi throughput, respectively, and are used to tradeoff their throughput and guarantee the fair coexistence. $|\mathcal{U}_s|$ and $|\mathcal{M}_s|$ represent the number of SUEs and the selected Wi-Fi users, respectively. $N$ is the number of spatial DoFs allocated to SUEs. $\bar{r}_s$ and $\bar{r}_w$ are the average data rate requirements for SUEs and the Wi-Fi users, respectively. The objective function in \eqref{op1} is to maximize the minimum weighted throughput of the small cell and the Wi-Fi systems. Constraints \eqref{op1_c2} and \eqref{op1_c3} represent the average data rate requirements for the SUEs and the Wi-Fi users, respectively. Constraint \eqref{op1_c4} means that the number of the selected SUEs should be less than or equal to the number of the DoFs allocated to the SUEs. Constraint \eqref{op1_c5} denotes the total available DoF constraint.

To solve problem \eqref{op1}, we set a new variable $z$ to replace the objective function in \eqref{op1}.
Then the problem is transformed into
\begin{eqnarray}\label{op2}
\small
\mathop {\max }\limits_{{\mathcal{U}_s},{\mathcal{M}_s},N}{z}
\end{eqnarray}
subject to~\eqref{op1_c2},~\eqref{op1_c3},~\eqref{op1_c4}, and
\begin{align}
\small
&z\ge e_sR_s,~z\ge e_wR_w. \label{op2_c1}\tag{\theequation a}
\end{align}
Since the number of spatial DoFs is an
integer, problem \eqref{op2} is a  mixed integer program problem.
To find the solution for problem \eqref{op2}, we first need to decide the number of spatial DoFs allocated to the serve the SUEs and mitigate the interference to the Wi-Fi users, respectively. After that, the number of the served SUEs and the Wi-Fi user selection for interference mitigation need to be decided. Therefore, two loops are required to find the solution for \eqref{op1}.

In the outer loop, a bisection method is applied to decide the spatial DoF allocation.
In the inner loop, the SBS selects the Wi-Fi users to maximize the average number of the active Wi-Fi users since the Wi-Fi throughput increases with the number of the served Wi-Fi users when it is not overloaded. Based on \cite{04_boyd}, the optimal solution for \eqref{op2} should satisfy the condition that $e_sR_s$ is equal or close to $e_wR_w$, which is used to judge whether the solution is reached.
The corresponding algorithm is summarized in Table I.
\begin{table}[!t]
\small
\caption{DoF allocation Algorithm}\label{alg1}\centering
\vspace{-0.4cm}
\begin{algorithm}[H]
\caption{}
\begin{algorithmic}[1]
\STATE Initialize a counter, $j$=0, the number of spatial DoF allocated to Wi-Fi users($D(j)=0$), $temp=big~value$,
$D_{\max}=N_T$, $D_{\min}=0$, $\Delta(j)=big~value$,
$\Delta p_m(j)=0 (the~increase~on~the~access~probability~if~Wi-Fi~user~m~is~selected),~\forall m\in\mathcal{M}$, $e_sR_s(j)$=0, $e_wR_w(j)$=0;
\WHILE {$|e_sR_s(j)-e_wR_w(j)|\le temp$}
    \STATE $temp=\Delta(j)$,  $D(j)=\lfloor \frac{D_{\max}+D_{\min}}{2}\rfloor$, where $\lfloor\cdot\rfloor$ is the least integer less than $\frac{D_{\max}+D_{\min}}{2}$, $N=N_t-D(j)-1$;
    \FOR{$m\in\mathcal{M}$}
    \STATE{Calculate the increase on access probability, $\Delta p_m(j)$, based on \eqref{eq4}, Lemma 2, and
    \eqref{eq41}, when SUE $m$ is selected for the interference mitigation};
    \ENDFOR
    \STATE{Sort Wi-Fi users by the value of $\Delta p_m(j)$ from big to small};
    \STATE{Select the first $D(j)$ Wi-Fi users to formulate a selected Wi-Fi user set, $\mathcal{M}_s(j)$}, calculate the average number of active Wi-Fi users;
    \STATE{Using the exhaustive search method to find the optimal number of SUEs, $K$, to maximize the
    small cell throughput based on \eqref{SB_th} with the available spatial DoF, $N$;}
    \STATE{Calculate SBS throughput, $R_s$, based on \eqref{SB_th}, and Wi-Fi throughput, $R_w$ based on
    \eqref{wf_throughput}};
    \IF {$e_sR_s(j)-e_wR_w(j)>0$}
        \STATE $j=j+1$, $D_{\max}=D(j)$;
    \ELSE
        \STATE $j=j+1$, $D_{\min}=D(j)$;
    \ENDIF
    \STATE $\Delta(j)=|e_sR_s(j)-e_wR_w(j)|$;
\ENDWHILE
\STATE{Output the DoF allocated to the Wi-Fi users, $D(j)$, selected Wi-Fi user set $\mathcal{M}_s(j)$, and the SBS throughput, $R_s(j)$, and the Wi-Fi throughput, $R_w(j)$};
\end{algorithmic}
\end{algorithm}
\end{table}

When selecting the Wi-Fi users in step 4 in Table I for interference mitigation, the increase on the channel access probability, $\Delta p_m(j)$, for each Wi-Fi user $m$ is first calculated based on \eqref{eq4}, \eqref{ga}, and \eqref{eq41} if Wi-Fi $m$ is selected. Then, the Wi-Fi users are sorted by the value of $\Delta p_m$ in decreasing order. The first $D(j)$ Wi-Fi users for the interference cancelation at the SBS are selected to maximize the number of the active Wi-Fi users. The exhaustive search method is applied to find the optimal number of the served SUEs to maximize the small cell throughput when the available spatial DoFs are $N=N_t-D(j)-1$ for the SUEs.


\section{Simulation Results}
Numerical simulation results are presented in this section to verify the above analysis and the effectiveness of the proposed schemes.
We assume that there is one SBS with 200 m coverage with one Wi-Fi AP in its coverage. The rest major simulation parameters are listed in Table II, where SIFS and DIFS are the short interframe space and the distributed interframe space in the DCF protocol, respectively.
\begin{table}[h]\center
\caption{Simulation parameters}\label{parameter}
{\small{
\begin{tabular}{c|c}
  \hline
  Parameters & Value\\
  \hline\hline
  \tabincell{c}{Number of antennas at the SBS} & $8$\\
  \hline
  \tabincell{c}{Wi-Fi packet\\ payload} & $12000 {\rm bits}$\\
  \hline
  \tabincell{c}{MAC header of \\Wi-Fi packet}& $192{\rm bits}$\\
  \hline
  \tabincell{c}{PHY header of \\Wi-Fi packet} & $224 {\rm bits}$\\
  \hline
  SIFS & $16{\rm\mu s}$ \\
  \hline
  DIFS & $34{\rm \mu s}$ \\
  \hline
  \tabincell{c}
  {Channel bit rate\\ for Wi-Fi} & $300{\rm Mbps}$ \\
  \hline
  \tabincell{c}
  {Wi-Fi backoff \\window size $W$} & $16{\rm \mu s}$\\
  \hline
  \tabincell{c}
  {Wi-Fi maximum \\backoff stage $L$} & 6\\
  \hline
  Wi-Fi slot time size, $\sigma$ & $20{\rm \mu s}$\\
  \hline
  Unlicensed bandwidth & $20{\rm MHz}$  \\
  \hline
  \tabincell{c}
  {Transmission power limit\\ on unlicensed channel} & $23{\rm mw}$ \\
  \hline
  AWGN noise power & $\left.{-174{\rm dBm}}\middle/{\rm Hz}\right.$\\
   \hline
  \tabincell{c}
  {Average data rate \\requirement of SUEs} & $10{\rm Mbps}$\\
  \hline
  \tabincell{c}
  {Average data rate \\requirement of Wi-Fi users} & $10{\rm Mbps}$\\
  \hline
\end{tabular}}}
\end{table}

\begin{figure}[!t]\centering
\begin{centering}
\includegraphics[width=0.45\textwidth]{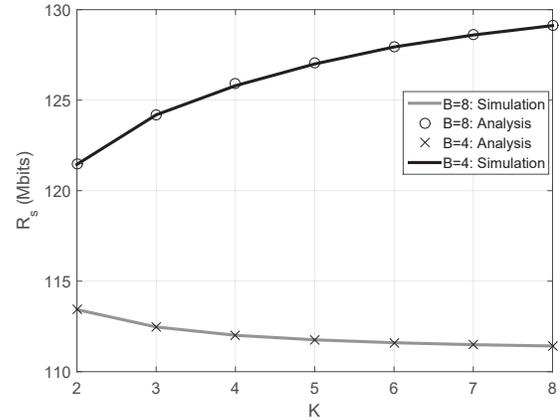}
\caption {Achievable small cell throughput when the number of served SUEs changes}\label{fig_2}
\end{centering}
\end{figure}

\begin{figure}[!t]\centering
\begin{centering}
\includegraphics[width=0.45\textwidth]{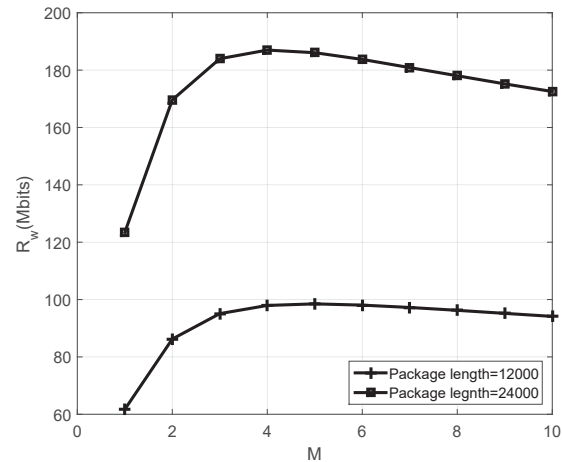}
\caption {Wi-Fi throughput when the number of served Wi-Fi users changes}\label{fig_3}
\end{centering}
\end{figure}

\subsection{Verification on the performance analysis}
First, we validate the theoretical claims in Section III through numerical simulation. In Fig. 2, we demonstrate the change on the small cell throughput with different numbers of SUEs served by the SBS. When the CSI feedback bits are 8, the small cell system throughput will increase with the number of served SUEs. On the other hand, if the number of feedback bits are 4, the maximum throughput is achieved when the number of served SUEs is 2. As we have explained after equation \eqref{SB_th} in Section III, when the number of CSI feedback bits is not enough, the small cell throughput may decrease with the increase on the number of the served SUEs due to the CSI quantization error. In this case, serving more SUEs is not helpful to the small cell throughput due to the limited CSI feedback bits. In the following simulation, we use 8 feedback bits to quantize the CSI between the SBS and SUEs. From the figure, the theoretical and simulation results are quite close.

Fig. 3 shows that the Wi-Fi system throughput varies with the number of the active Wi-Fi users. We assume the Wi-Fi users have the same type of traffic. Therefore, the number of the Wi-Fi users can be used to evaluate the Wi-Fi traffic load in our work. We can observe that the Wi-Fi throughput is convex with respect to the number of the active Wi-Fi users. When the number of the Wi-Fi users is fewer than 5, the Wi-Fi throughput will increase with the number of the Wi-Fi users. On the other hand, if the number of the Wi-Fi users is larger than 5, the Wi-Fi throughput will decrease with the number of Wi-Fi users.

\subsection{Performance of Algorithm 1}
In this section, we assume that the number of the Wi-Fi users is fewer than 5. To verify the performance of Algorithm 1, we demonstrate the achievable throughput of the SBS and the Wi-Fi system in Table III when different weight factors are assigned to them. As we can observe, when the weight factor for the Wi-Fi throughput is bigger than that for the small cell throughput, most spatial DoFs will be allocated to the Wi-Fi users via the algorithm. On the other hand, as the the weight factor for the SBS throughput increases, the achievable throughput at the SBS also increases since more spatial DoFs will be available to the SUEs. Therefore, the tradeoff between the small cell throughput and Wi-Fi throughput can be obtained via Algorithm 1 when the Wi-Fi system is not overloaded. Notice that the LTE utilizes the resource more sufficient than the Wi-Fi system. Therefore, more spatial DoFs have to be allocated to the Wi-Fi users. To explain this point more clearly, we use one snapshot to show the change on SBS and Wi-F throughput with the spatial DoF allocation in Fig. 4. From the figure, the small cell throughput increases when its allocated spatial DoF increases while the achievable throughput at Wi-Fi decreases. Moreover, the achievable throughput at the SBS is always higher than the Wi-Fi throughput since the efficient utilization on the unlicensed spectrum by LTE-U.
\begin{table}[h]\center
\caption{Simulation parameters}\label{parameter}
\small
\begin{tabular}{c|c|c|c}
  \hline
  Weight factor & \tabincell{c}{Small cell\\ throughput \\(Mbits)} & \tabincell{c}{Wi-Fi \\throughput \\(Mbits)} & \tabincell{c}{Spatial DoF \\allocation}\\
  \hline\hline
   ($0.1,0.9$)& $130.57$ & $61.71$ & ($6,1$)\\
  \hline
  ($0.15,0.85$)& $130.57$ & $61.71$ &($6,1$) \\
  \hline
   ($0.2,0.8$)& $130.57$  & $61.71$ & ($6,1$) \\
  \hline
  ($0.25,0.75$)& $130.57$  & $61.71$ & ($6,1$) \\
  \hline
  ($0.3,0.7$)& $130.57$  & $61.71$ & ($6,1$) \\
  \hline
  ($0.35,0.65$)& $129.6$  & $86$ & ($5,2$) \\
  \hline
  ($0.4,0.6$)& $129.6$  & $86$ & ($5,2$) \\
  \hline
  ($0.45,0.55$)& $125$  & $98$ & ($2,5$) \\
  \hline
  ($0.5,0.5$)& $125$  & $98$ & ($2,5$) \\
  \hline
\end{tabular}
\end{table}

\begin{figure}[!t]\centering
\begin{centering}
\includegraphics[width=0.45\textwidth]{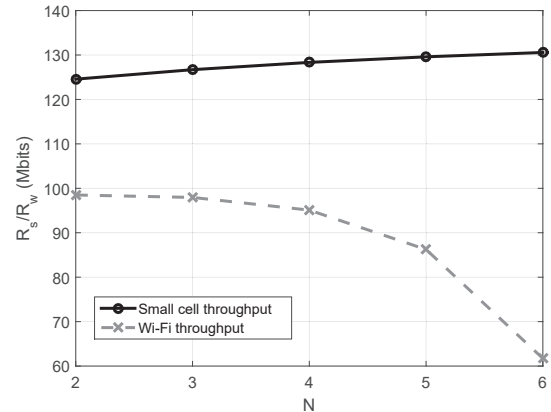}
\vspace{-2cm}
\caption {Small cell and Wi-Fi throughput v.s. spatial DoF allocation}\label{fig_4}
\end{centering}
\end{figure}

\section{Conclusions}
In the paper, we have applied the ZFBF technique to the SBS in the LTE-U system to realize the spectrum reuse on the unlicensed band with Wi-Fi system. The spatial DoF allocation scheme is developed to tradeoff the small cell and the Wi-Fi throughput after the analysis on small cell and Wi-Fi throughput. Through the simulation, the theoretical analysis and the effectiveness of the proposed scheme have been verified.

\section{Acknowledgement}
This work was supported in part by National Natural Science Foundation Program of China under Grant No. 61771429 No. 61301143, and No. 61771428, and Natural Science Foundation Program of Zhejiang Province under Grant No. LY14F010002.

\end{document}